\documentclass[12pt]{article}
\usepackage[english]{babel}
\usepackage[T1]{fontenc}
\usepackage{amsmath,amsfonts,graphicx,color,braket,cite}
\usepackage{ulem}
\newcommand{\no}{\nonumber }
\newcommand{\beeq}{\begin{equation}}
\newcommand{\eeq}{\end{equation}}
\newcommand{\bee}{\begin{eqnarray}}
\newcommand{\ee}{\end{eqnarray}}
\newcommand{\been}{\begin{eqnarray*}}
\newcommand{\een}{\end{eqnarray*}}

\begin{document}
\begin{titlepage}
\begin{flushright}
TIT/HEP-667\\
May, 2018
\end{flushright}
\vspace{0.5cm}
\begin{center}
{\large \bf
Massive ODE/IM Correspondence and Non-linear Integral Equations for $A_r^{(1)}$-type modified Affine Toda Field Equations
}
\lineskip .75em
\vskip 2.5cm
{\large  Katsushi Ito and Hongfei Shu }
\vskip 2.5em
 {\normalsize\it Department of Physics,\\
Tokyo Institute of Technology\\
Tokyo, 152-8551, Japan} 
\vskip 3.0em
\end{center}

\begin{abstract}
The massive ODE/IM correspondence is a relation between the linear problem associated with modified affine Toda field equations and two-dimensional massive integrable models. 
We study the massive ODE/IM correspondence for the $A_r^{(1)}$-type modified affine Toda field equations.
Based on the $\psi$-system satisfied by the solutions of the linear problem, we derive the Bethe ansatz equations and determine the asymptotic behavior of the $Q$-functions for large value of the spectral parameter. 
We derive the non-linear integral equations for the $Q$-functions from the Bethe ansatz equations.
We compute the effective central charge in the UV limit, which is identified with the one of the non-unitary $WA_r$ minimal models when the solution has trivial monodromy around the origin of the complex plane.

 \end{abstract}
\end{titlepage}
\baselineskip=0.7cm
\section{Introduction} 
The ODE/IM correspondence proposed in \cite{Dorey:1998pt,Bazhanov:1998wj} describes a relation between the spectral analysis of ordinary differential equation (ODE), and the ``functional relations'' approach to 2d quantum integrable model (IM). This correspondence provides an example of nontrivial relations between classical and quantum integrable models, which  plays an important role in studying strong coupling physics of supersymmetric gauge theories \cite{Alday:2009dv,Alday:2010vh,Hatsuda:2010cc,Maldacena:2010kp,Hatsuda:2010vr,Hatsuda:2011ke,Hatsuda:2011jn,Hatsuda:2012pb,Hatsuda:2014vra,Gaiotto:2014bza,Ito:2017ypt}.
In order to understand this non-trivial correspondence, it is important to identify the integrable models from the ODEs.

A basic strategy of identifying the quantum integrable model is to obtain the functional relations from the ODE, from which we can study the energy spectrum of the integrable model.
In particular, starting from the solutions to the ODE, one can derive the functional relations, such as the T-Q relations, the Bethe ansatz equations and the T-/Y-systems (see \cite{Kuniba:2010ir} for a review). Furthermore, the functional relations are converted to the non-linear integral equations.
In the UV and IR limit, one obtains the effective central charge of the quantum integrable model.

From the ODEs, one obtains the massless integrable models or conformal field theories \cite{Dorey:1999pv,Suzuki:1999hu,Dorey:2000ma,Suzuki:2000fc,Suzuki:2000gi,Dorey:2001uw,Bazhanov:2003ni,Dorey:2006an,Dorey:2007zx,Feigin:2007mr,Sun:2012xw,Masoero:2010is,Suzuki:2015vwa,Masoero:2015lga,Masoero:2015rcz,Negro:2017xwc}.
Recently, the ODE/IM correspondence has been generalized to the case of massive integrable models \cite{Lukyanov:2010rn, Dorey:2012bx,Ito:2013aea,Adamopoulou:2014fca,Ito:2015nla,Locke-Thesis,Negro:2016yuu}\footnote{See \cite{Lukyanov:2011wd, Lukyanov:2012wq, Lukyanov:2013wra, Bazhanov:2013cua,Bazhanov:2013oya, Bazhanov:2014joa, Bazhanov:2016glt,Vicedo:2017cge,Bazhanov:2017xky} for the correspondences to the integrable non-linear sigma models and the affine Gaudin models.}. 
In this massive ODE/IM correspondence, the ODE is replaced to the linear problem associated with the modified 
affine Toda field equation based on the Langlands dual of an affine Lie algebra $\hat{\mathfrak g}$. Taking the light-cone limit, the linear problem reduces to the ODE, from which one obtains the Bethe ansatz equations associated with the affine Lie algebra $\hat{\mathfrak{g}}$. 

A powerful method to study the Bethe
ansatz equations on a space of finite length is %known as 
the non-linear integral equations (NLIEs) \cite{
Klummpe:1991vs,Klumper-Pearce,Destri:1992ey}, which are more easily to evaluate and can be solved numerically. The NLIE for the quantum sine-Gordon model was studied in \cite{Destri:1992ey,Destri:1994bv,Destri:1997yz, Feverati:1998dt, Feverati:1999sr}. It has been generalized to the complex affine Toda models associated to simply laced algebras \cite{ZinnJustin:1997at}.
In \cite{Dorey:2000ma} the NLIEs for the $A_r$-type have been derived from the $(r+1)$-th order ODE, where the NLIEs are the massless limit of those in \cite{ZinnJustin:1997at}. 
More recently, the massive NLIE of the quantum sine-Gordon model has been directly derived from the $A_1^{(1)}$-type modified Toda field  equation. The purpose of this paper is to construct the massive NLIEs for modified affine Toda field equations
and to identify the quantum integrable models in the UV limit. In the present work we will discuss the $A_r^{(1)}$-type modified affine Toda field equations as a non-trivial generalization.

This paper is organized as follows:
In sect.~\ref{sec:MATF-Rew}, we discuss the $A_r^{(1)}$-type modified affine Toda field equations and their associated linear problem. We study the asymptotics of the solution to the linear problem, and introduce the Q-functions. In sect. \ref{sec:BAEs}, we derive the Bethe ansatz equations for Q-functions from the $\psi$-system satisfied by the solutions to the linear problems.
In sect.~\ref{sec:Reduction-Q}, we study the analytic properties of the Q-functions  by taking the light-cone limit of the linear problem.
In sec.~\ref{sec:NLIEs-Ar}, we derive the non-linear integral equations from the Bethe ansatz equations of Q-functions. We then study the UV limit of the associated massive integrable model, and obtain the effective central charge. Sec.~\ref{sec:conclusions and discussion} contains conclusions and discussions. In appendix \ref{sec:massive-IM}, the detailed form of the NLIEs for the $A_r$-type complex affine Toda model is presented.

%%%%%%%%%%%%%%%%%%%%%%%%%%%%%%%%%%%%%%%%%%%%%%%%%%%%%%%%%%%%%%%%%%%%%%%%%%%%%%%%%%%%%%%%%%%%%%%%%%%%%%%%%%%%%%%%%%%%%%%%%%%%%%%%%%%%%%%%%%%%

\section{Modified affine Toda field equation and Q-function}~\label{sec:MATF-Rew}
In this section we introduce the linear problem associated with the $A_r^{(1)}$-type modified affine Toda field 
equations and study the asymptotic behavior of the solutions \cite{Ito:2013aea,Ito:2015nla}.

\subsection{Lie algebra $A_r$}
We begin with some definitions of the Lie algebra
$\mathfrak{g}=A_r$  with rank $r$.
The generators are denoted by $\{H^a,E_\alpha\} (a=1,\cdots, r, \alpha\in \Delta$). Here $\Delta$ is a set of roots, normalized such that 
the squared length is $2$.
The simple root $\alpha_a$  and the fundamental weights $\omega_a$ ($a=1,\cdots,r$) of $\mathfrak{g}$ satisfy 
$\alpha_a\cdot \alpha_b=C_{ab}$,
$\alpha_a\cdot \omega_b=\delta_{ab}$.
Here $C_{ab}=2\delta_{ab}-\delta_{a,b+1}-\delta_{a,b-1}$ is the Cartan matrix of $A_{r}$.
The affine Lie algebra $\hat{\mathfrak{g} }=A_r^{(1)}$ is given by adding the root $\alpha_0=-\theta$ to $\mathfrak{g}$, where $\theta:=\alpha_1+\dots+\alpha_r$ is the highest root.
The Weyl vector $\rho$ is defined by $\rho=\omega_1+\dots+\omega_r$.

Let $V^{(a)}$ be the basic $\mathfrak{g}-$module associated with the highest weight $\omega_a$. We denote the orthonormal basis of $V^{(a)}$ as ${\bf e}_j^{(a)}$ ($j=1,\cdots,\mbox{dim}V^{(a)})$, which are the eigenvectors of the Cartan generator $H^b$ with eigenvalue $(h^{(a)}_j)^b$. $h^{(a)}_j$ is the weight vector of $V^{(a)}$.
For example, $V^{(1)}$ is the $(r+1)$-dimensional fundamental representation with the highest weight $\omega_1$, whose matrix representation is given by 
\beeq
	E_{\alpha_{0}}=e_{r+1,1},\quad E_{\alpha_{a}}=e_{a,a+1},\quad a=1,\cdots,r
\eeq
where $e_{a,b}$ denotes the matrix with non-zero components $(e_{a,b})_{cd}=\delta_{ac}\delta_{bd}$.
$E_{-\alpha_a}$, $E_{-\alpha_0}$ and $H^a$ ($a=1,\dots,r$) are defined by $E_{-\alpha_{0}}=E_{\alpha_{0}}^{\top}$ and
$E_{-\alpha_{a}}=E_{\alpha_{a}}^{\top}$ and $\alpha_a\cdot H=[E_{\alpha_b},E_{-\alpha_b}]$.
The weight vectors $h_1^{(1)},\cdots,h_{r+1}^{(1)}$ of  $V^{(1)}$ satisfy
$h^{(1)}_1=\omega_1$ and 
\beeq
	h_{a}^{(1)}-h_{a+1}^{(1)}=\alpha_{a},\quad \sum_{a=1}^{r+1}h_a^{(1)}=0.
\eeq

\subsection{$A_r^{(1)}$-type modified affine Toda field equation}
Let $\phi=(\phi_1,\phi_2,\cdots ,\phi_r)$ be the $r$-component scalar field on the $(z,\bar{z})$ complex plane.
The $A_r^{(1)}$-type modified affine Toda field equations are defined by
	\beeq\label{eq:MATFE}
		\partial_z{\partial}_{\bar{z}}\phi - \frac{m^{2}}{\beta}\left[\sum_{a=1}^{r}%n_{i}
        \alpha_{a}\exp[\beta\alpha_{a}\cdot\phi]+p(z)\bar{p}(\bar{z})%n_{0}
        \alpha_{0}\exp[\beta\alpha_{0}\cdot\phi]\right]=0
	\eeq
	where $\beta$ is a dimensionless coupling parameter and $m$ a mass parameter. 
$p(z)$ and $\bar{p}(\bar{z})$ in (\ref{eq:MATFE}) are defined by
	\beeq\label{eq:p(z)}
		p(z)=z^{hM}-s^{hM},~~\bar{p}(\bar{z})=\bar{z}^{hM}-\bar{s}^{hM}
	\eeq 
for a complex parameter $s$ and a positive real number $M>\frac{1}{h-1}$.	
Here $h=r+1$ is the Coxeter number of $A_r$.
Eq.(\ref{eq:MATFE}) is
regarded as the compatibility condition of the linear problem
    \begin{equation}\label{eq:LP-2}
		(\partial_z+{ A}_z)\Psi=0,~~({\partial}_{\bar{z}}+{ A}_{\bar{z}})\Psi=0,%\quad {\bf A}=A_zdz+{A}_{\bar{z}}d\bar{z}.
	\end{equation}
    where $A_z$ and $A_{\bar{z}}$ are defined by
	\begin{eqnarray}\label{eq:Flat-Connection}
		&&A_z=\frac{\beta}{2}\partial_z\phi\cdot H+me^{\lambda}\left\{\sum_{a=1}^{r}%\sqrt{n_{i}^{\vee}}
        E_{\alpha_{a}}\exp(\frac{\beta}{2}\alpha_{a}\cdot\phi)+p(z)%\sqrt{n_{0}^{\vee}}
        E_{\alpha_{0}}\exp(\frac{\beta}{2}\alpha_{0}\cdot\phi)\right\},
		\\
		&&
		{A}_{\bar{z}}=-\frac{\beta}{2}{\partial}_{\bar{z}}\phi\cdot H + me^{-\lambda}\left\{\sum_{a=1}^{r}%\sqrt{n_{i}^{\vee}}
        E_{-\alpha_{a}}\exp(\frac{\beta}{2}\alpha_{a}\cdot\phi)+\bar{p}(\bar{z})%\sqrt{n_{0}^{\vee}}
        E_{-\alpha_{0}}\exp(\frac{\beta}{2}\alpha_{0}\cdot\phi)\right\}.\nonumber
	\end{eqnarray}
Here we have introduced a spectral parameter $\lambda$.
	
\subsection{Asymptotic behaviors and symmetries}
We study a class of solutions $\phi(z,\bar{z})$ of eq.(\ref{eq:MATFE})  
satisfying the periodic condition
$\phi(|z|,\theta+\frac{2\pi}{hM})=\phi(|z|,\theta)$  where we have introduced the polar coordinate $z=|z|e^{i\theta}$. 
They are also required to satisfy the boundary conditions at
$|z|=\infty$ and $0$, which are given by %infinity and the origin 
		\begin{align}
			\phi(z,\bar{z})&=\frac{M\rho}{\beta}\log(z\bar{z})+\cdots~~~(|z|\to \infty),\\
		\phi(z,\bar{z})&=g\log(z\bar{z})+\phi^{(0)}(g)+\gamma(z,\bar{z},g)+\sum_{a=0}^{r}\frac{C_{a}(g)}{(c_{a}(g)+1)^{2}}(z\bar{z})^{c_{a}(g)+1}+\cdots\label{eq:phi-expansion}~~~(|z|\to 0),\no\\
		\end{align}
		where $g$ is a $r$-component vector satisfying $\beta\alpha_m\cdot g+1>0$ for $m=0,1,\cdots,r$.  $\phi^{(0)}(g)$ is a constant vector. $\gamma(z,\bar{z},g)$ is given by 
        \bee
        \gamma(z,\bar{z},g)=\sum_{k=1}^\infty \gamma_k(g)(z^{hMk}+\bar{z}^{hM k})
        \ee
with some coefficients $\gamma_k(g)$.
Other coefficients are given by $c_{a}=\beta\alpha_{a}\cdot g$, $C_{a}=-\frac{m^{2}}{\beta}\alpha_{a}e^{\beta\alpha_{a}\cdot\phi^{(0)}(g)}$ and 
$C_{0}=\frac{m^{2}}{\beta}(s\bar{s})^{hM}\alpha_{0}e^{\beta\alpha_{0}\cdot\phi^{(0)}(g)}$.

  In the following, we regard $z$ and $\bar{z}$ as independent variables. The linear problem (\ref{eq:LP-2}) are invariant under the Symanzik roation $\hat{\Omega}_k$ with integer $k$, which is defined by
  \beeq
		\hat{\Omega}_k=
		\left\{
		\begin{array}{ccc}
			z\to ze^{\frac{2\pi ki}{hM}}\\
			s\to se^{\frac{2\pi ki}{hM}}\\
			\lambda\to\lambda-\frac{2\pi ki}{hM}
		\end{array}
		\right.,
	\eeq
    which acts on the function with arguments $z,\bar{z}$ and $\lambda$. 
It is also invariant under the transformation $\hat{\Pi}$ defined by
    \bee
\hat{\Pi}:
		\left\{
		\begin{array}{ccc}
			\lambda\to \lambda-\frac{2\pi i}{h}\\
			({A}_z, A_{\bar{z}})\to S({A}_z, A_{\bar{z}})S^{-1}\\
			\Psi\to S\Psi
		\end{array}
		\right.,\quad S=\exp\left(\frac{2\pi i}{h}\rho^{\vee}\cdot H\right).
\ee

\subsection{Asymptotic behaviors of solutions to linear problem}
We now study the solution $\Psi$ to the linear problem (\ref{eq:LP-2}).
In the large $|z|$ region, we obtain the asymptotic solution of the linear problem by using the WKB analysis{\cite{Ito:2015nla}}. In the module $V^{(a)}$, the fastest decaying asymptotic solution along the positive part of the real axis for large $|z|$ is
	\beeq\label{eq:Asy_LP}
		\Xi^{(a)}(|z|,\theta)\sim C^{(a)} e^{-i\theta M\rho\cdot H}\exp\left(-\mu^{(a)}\frac{2|z|^{M+1}}{M+1}m\cosh[\lambda+i\theta(M+1)]\right)e^{-i\theta M\rho\cdot H}\hat{\mu}^{(a)},
	\eeq
	where $C^{(a)}$ is a constant. $\hat{\mu}^{(a)}$ is the eigenvectors of $\Lambda=E_{\alpha_{0}}+\sum_{b=1}^{r}E_{\alpha_{b}}$ with the largest real eigenvalue $\mu^{(a)}$ in $V^{(a)}$. This WKB solution is valid in the range of $|\theta|<\frac{(h+1)\pi}{h(M+1)}$, and 
 is the subdominant one in the Stokes sector ${\cal S}_0$ where the sector ${\cal S}_k$ ($k\in {\bf Z}$) is defined by
	\beeq\label{eq:Stokes-sec-0}
		{\cal S}_{k}:~|\theta-\frac{2\pi k}{h(M+1)}|<\frac{\pi}{h(M+1)}.
	\eeq
 There are $r+1$ independent solutions in each Stokes sector. 
 We denote the subdominant solution in ${\cal S}_0$ as $s_0^{(a)}$, which is uniquely defined. The asymptotic behavior for large $|z|$ of $s_0^{(a)}$ is given by $\Xi^{(a)}$. We introduce the subdominant solution 
in ${\cal S}_k$ as $s_k^{(a)}$, which 
is obtained from $s_0^{(a)}$ by the Symanzik rotation 
	\begin{eqnarray}\label{eq:small-solution}
		s_k^{(a)}=\hat{\Omega}_{-k}s_0^{(a)}.
	\end{eqnarray}
The asymptotics of $s_k^{(a)}$ is determined by  $\hat{\Omega}_{-k}\Xi^{(a)}$.

	In the small $|z|$ region, the asymptotic solution is given by 
	\begin{eqnarray}\label{eq:basis-Ar}
		{\cal X}_i^{(a)}=B_i^{(a)}(g) e^{-(\lambda+i\theta)\beta g\cdot h_{i}^{(a)}}{\bf e}_i^{(a)}+{\cal O}(|z|), \quad i=1,\cdots,\dim V^{(a)},
	\end{eqnarray}
   where $B_i^{(a)}(g)$ is a constant. ${\cal X}_i^{(a)}$ ($i=1,\cdots,\mbox{dim}V^{(a)}$) form an orthonormal basis of the solution to linear problem. They are invariant under the Symanzik rotation.

We expand $s_0^{(a)}$ in terms of the basis ${\cal X}_i^{(a)}$
	\beeq\label{eq:Q-function}
		s_{0}^{(a)}(z,\lambda)=\sum_{i=1}^{\dim V^{(a)}}Q_{i}^{(a)}(\lambda){\cal X}_{i}^{(a)}(z).
	\eeq
  As we will see later, the coefficients become the Q-functions of the integrable model. From $\hat{\Omega}_{-1}\hat{\Pi}s_0=s_0$, 
  one finds a quasi-periodic condition
	\beeq\label{eq:qusi-period}
		Q_{i}^{(a)}(\lambda-\frac{2\pi i}{hM}(M+1))=\exp\left(-\frac{2\pi i}{h}(\rho^{\vee}+\beta g)\cdot h_{i}^{(a)}\right)Q_{i}^{(a)}(\lambda).
	\eeq

%%%%%%%%%%%%%%%%%%%%%%%%%%%%%%%%%%%%%%%%%%%%

%%%%%%%%%%%%%%%%%%%%%%%%%%%%%%%%%%%%%%%%%%%%

\section{Bethe ansatz equations}\label{sec:BAEs}
In this section, we derive the Bethe ansatz equations for $Q_1^{(a)}(\lambda)$ by using the $\psi-$system \cite{Dorey:2006an, Sun:2012xw, Masoero:2015lga,Ito:2015nla}.
We also derive the T-Q relations, which are obtained from the relations among the determinants of the Q-functions. We then discuss their relations to the Bethe ansatz equations.

\subsection{$\psi$-system and Bethe ansatz equations}
The subdominant solutions $\Psi^{(a)}$ to the linear problem in a different module $V^{(a)}$ are not independent of each other. 
They obey the relations called the $\psi$-system, which is defined by the inclusion maps $\iota$ from the antisymmetric representation $V^{(a)}\wedge V^{(a)}$ to the representation $V^{(a-1)}\otimes V^{(a+1)}$ \cite{Sun:2012xw}:
\bee\label{eq:psi-system-map}
\iota\left(V^{(a)}\wedge V^{(a)}\right)=V^{(a-1)}\otimes V^{(a+1)},\quad a=1,\cdots,r,
\ee
where $V^{(0)}=V^{(r+1)}=\mathbb{C}$.
By comparing the asymptotic behaviors of the solutions of the linear problem for large $|z|$, one finds
\bee\label{eq:psi-system}
\iota\left(\Psi^{(a)}_{[-\frac{1}{2}]}\wedge \Psi^{(a)}_{[\frac{1}{2}]}\right)=\Psi^{(a-1)}\otimes \Psi^{(a+1)},
\ee
where $\Psi^{(0)}=1=\Psi^{(r+1)}$. The subscript ${[k]}$ implies that $f_{[k]}(z,\lambda):=\hat{\Omega}_kf(z,\lambda)$ ($k\in {\bf Z}$ or ${\bf Z}+{1\over2}$). Substituting (\ref{eq:Q-function}) to the $\psi$-system (\ref{eq:psi-system}) and comparing the top components in both sides, we obtain
\bee\label{eq:first-component-psi-system}
Q_{1,[-\frac{1}{2}]}^{(a)}Q_{2,[\frac{1}{2}]}^{(a)}-Q_{1,[\frac{1}{2}]}^{(a)}Q_{2,[-\frac{1}{2}]}^{(a)}&=Q_{1}^{(a-1)}Q_{1}^{(a+1)},
\ee
where $Q^{(0)}_{1,2}=Q_{1,2}^{(r+1)}=1$.
Letting the zeros of $Q^{(a)}_1(\lambda)$ be $\lambda^{(a)}_{j}$ ($j\in \mathbb{Z}$), one obtains the equations
\beeq\label{eq:BAE-psi}
-1=\prod_{b=1}^{r}\frac{Q_{1}^{(b)}(\lambda_{j}^{(a)}+C_{ab}\frac{\pi i}{hM})}{Q_{1}^{(b)}(\lambda_{j}^{(a)}-C_{ab}\frac{\pi i}{hM})},
\eeq
where $C_{ab}$ is the Cartan matrix of $A_r$. This set of equations (\ref{eq:BAE-psi}) provides the Bethe ansatz equations for the massive integrable models \cite{Adamopoulou:2014fca,Ito:2015nla}. In section \ref{sec:NLIEs-Ar}, we will study the analytic properties of $Q^{(a)}_i(\lambda)$.% to justify the mean of massive version. 

\subsection{T-Q relations and Bethe ansatz equations}\label{subsec:BAEs-Ar}
We consider the solutions of the linear problem in the whole complex plane. We introduce a {skew-symmetric} product of the solutions $s^{(1)}_i$ in the fundamental $(r+1)$-matrix representation $V^{(1)}${, which is defined by}
\beeq
\braket{s^{(1)}_{i_1}, s^{(1)}_{i_2},\cdots,s^{(1)}_{i_{r+1}}}{\equiv}\det\left(s^{(1)}_{i_1}, s^{(1)}_{i_2},\cdots,s^{(1)}_{i_{r+1}}\right){.}
\eeq
 This is independent of $z$.
Using the asymptotic behaviors of $s_{i_k}^{(1)}$, one can compute this product explicitly. 
The normalization constant $C^{(1)}$ in (\ref{eq:Asy_LP}) is determined to 
satisfy
	\begin{equation}\label{eq:norm}
	\braket{s_{i}^{(1)},s^{(1)}_{i+1},\cdots,s^{(1)}_{i+r}}=1.
	\end{equation}

Note that $V^{(p)}$ for $p=1,\dots, r+1$ is obtained as the exterior product of $V^{(1)}$, i.e. $\bigwedge^pV^{(1)}$. 
In the exterior product $s_{i_1}^{(1)}\wedge s_{i_2}^{(1)}\wedge \cdots \wedge s_{i_p}^{(1)}\in\bigwedge^pV^{(1)}$, the coefficient of the highest weight vector ${\cal X}_1^{(1)}\wedge {\cal X}_2^{(1)}\wedge \cdots\wedge {\cal X}_p^{(1)}$ is expressed as the determinant of the $p\times p$ matrix,
whose $(k,\ell)$ element is given by $Q^{(1)}_{k[-i_\ell]}(\lambda)${. We define } 
\begin{equation}\label{eq:Wronskian}
       W^{(p)}_{i_1,i_2,\ldots, i_p}(\lambda){\equiv}
\det Q^{(1)}_{k[-i_\ell]}(\lambda).  %\left(\begin{array}{cccc}
%Q_{1}(\lambda+i_{1}\frac{2\pi i}{hM}) & Q_{1}(\lambda+i_{2}\frac{2\pi i}{hM}) & 
%\cdots & Q_{1}(\lambda+i_{p}\frac{2\pi i}{hM})\\
%Q_{2}(\lambda+i_{1}\frac{2\pi i}{hM}) & Q_{2}(\lambda+i_{2}\frac{2\pi i}{hM}) & 
%\cdots & Q_{2}(\lambda+i_{p}\frac{2\pi i}{hM})\\
%\vdots & \vdots &  & \vdots\\
%Q_{p}(\lambda+i_{1}\frac{2\pi i}{hM}) & Q_{p}(\lambda+i_{2}\frac{2\pi i}{hM}) & 
%\cdots & Q_{p}(\lambda+i_{p}\frac{2\pi i}{hM})
%\end{array}\right).
\end{equation}
For $p=1$, we have $W^{(1)}_{i_1}(\lambda)=Q_1^{(1)}(\lambda+i_1\frac{2\pi i}{hM})$. For $p=2$ with $i_1={1\over2}$ and $i_2=-{1\over2}$, $W^{(2)}_{-\frac{1}{2},\frac{1}{2}}(\lambda)=Q_1^{(2)}(\lambda)$, where we used (\ref{eq:first-component-psi-system}).
In general, for $p=1,\dots,r$, we express $Q^{(p)}_1(\lambda)$ as
\begin{align}\label{eq:Q1a-Qa}
Q^{(p)}_1(\lambda)=W_{-\frac{p-1}{2},1-\frac{p-1}{2},\cdots,\frac{p-1}{2}}^{(p)}(\lambda).
\end{align}
For $p=r+1$, using the normalization condition (\ref{eq:norm}), we find
	\begin{eqnarray}\label{eq:Wron-norm}
W^{(r+1)}_{i_1,i_1+1,\cdots,i_{1}+r}=\left[{\det}\left({\cal X}^{(1)}_1,\cdots,{\cal X}^{(1)}_{r+1}\right)\right]^{-1},
\end{eqnarray}
which imposes the constraints for the Q-functions.

%%%%%%%%%%%%%%%%%%%%%%%%%%%%%%%%%%%%%%%%%%%%%%%%%%%%%%%%%%%%%%%%%%%%%%%%%%%%%%%%%%%%%%%%%%%%%%%%%%%%%%%%%%%%%%%%%%%%%%%%%%%%%%%%%%%%%%%%%%%%

Note that the determinants (\ref{eq:Wronskian}) satisfy the Pl\"ucker relations
\begin{eqnarray}\label{eq:Plucker-1-2-0}
	W_{i_{0},i_{2},\cdots,i_{p-1}}^{(p-1)}W_{i_{1},i_{2},\cdots,i_{p}}^{(p)}-W_{i_{1},i_{2},\cdots,i_{p-1}}^{(p-1)}W_{i_{0},i_{2},\cdots,i_{p}}^{(p)}+W_{
    %, 
    {i_{2},\cdots,i_{p-1},i_{p}}}^{(p-1)}W_{i_{0},i_{1},\cdots,i_{p-1}}^{(p)}=0.
\end{eqnarray}
For $(i_0,i_1,\cdots,i_{p-1},i_p)=(0,1,\cdots,p-1,p)$, (\ref{eq:Plucker-1-2-0}) leads to 
\beeq
\frac{W_{0,2,\cdots, p}^{(p)}}{W_{1}^{(p)}}=\sum_{m=0}^{p-1}\frac{W_{2,\cdots,m+1}^{(m)}W_{0,\cdots,m}^{(m+1)}}{W_{1,\cdots,m}^{(m)}W_{1,\cdots,m+1}^{(m+1)}}\label{eq:T-Q-01}
\eeq
where $W_k^{(0)}=1$ for any $k$. Setting $p=r+1$ and shifting the spectral parameter by $\lambda\to \lambda-\frac{2\pi i}{hM}$, (\ref{eq:T-Q-01}) leads to
\beeq\label{eq:T-Q-1}
	W_{-1,1,\cdots,r}^{(r+1)}\prod_{j=0}^{r}W_{0,\cdots,j-1}^{(j)}=\sum_{m=0}^{r}\left(\prod_{j=0}^{m-1}W_{0,\cdots,j-1}^{(j)}\right)W_{1,\cdots,m}^{(m)}W_{-1,\cdots,m-1}^{(m+1)}\left(\prod_{j=m+2}^{r+1}W_{0,\cdots,j-1}^{(j)}\right).
\eeq
This is the T-Q relation in \cite{Dorey:2000ma}, where 
$W_{-1,1,\cdots,r}^{(r+1)}$ is regarded as the T-function.

Let $\Lambda_j^{(a)} (j\in \mathbb{Z})$ be the zeros of $W^{(a)}_{0,1,\cdots,a-1}(\lambda)$, (\ref{eq:T-Q-1}) leads to
\beeq\label{eq:Wron-BAEs}
	-1=\left. \frac{W_{0,\cdots,a-2}^{(a-1)}W_{1,\cdots,a}^{(a)}W_{-1,\cdots,a-1}^{(a+1)}}{W_{1,\cdots,a-1}^{(a-1)}W_{-1,\cdots,a-2}^{(a)}W_{0,\cdots,a}^{(a+1)}}\right|_{\lambda=\Lambda_j^{(a)}},\quad a=1,\cdots,r.
\eeq
From (\ref{eq:Q1a-Qa}) and identifying $\lambda_j^{(a)}=\Lambda_j^{(a)}+\frac{a-1}{2}\frac{2\pi i}{hM}$,
(\ref{eq:Wron-BAEs}) becomes the Bethe ansatz equations (\ref{eq:BAE-psi}).

As shown in (\ref{eq:Q1a-Qa}), $Q^{(a)}_1(\lambda)$ is expressed as the Wronskian of $Q^{(1)}_i(\lambda)$. One finds a similar Wronskian representation for $Q_2^{(a)}(\lambda)$.
This follows by the identity between the determinants
\beeq
\Delta^{(a+1)}\Delta^{(a-1)}[a,a+1|1,a+1]=\Delta^{(a)}[a+1|a+1]\Delta^{(a)}[a|1]-\Delta^{(a)}[a+1|1]\Delta^{(a)}[a|a+1].\label{eq:Jacobi-ID}
\eeq
Here $\Delta^{(a+1)}$ is the determinant of an $(a+1)\times (a+1)$-matrix. $\Delta^{(a)}[p_1|q_1]$ is the determinant of the $a\times a$ matrix with $p_{1}$ row and $q_{1}$ column removed from the matrix of $\Delta^{(a+1)}$. $\Delta^{(a-1)}[p_1,p_2|q_1,q_2]$ is the determinant of the $(a-1)\times (a-1)$ matrix obtained by
removing  $p_{1,2}$ rows and $q_{1,2}$ columns of the matrix of $\Delta^{(a+1)}$.
Using (\ref{eq:Jacobi-ID}), we find
\bee\label{eq:Q2a-Qa}
Q_2^{(a)}(\lambda)=W^{(a+1)}_{-\frac{a-1}{2},1-\frac{a-1}{2},\cdots,\frac{a+1}{2}}[a|a+1](\lambda).
\ee

We have mentioned that in  (\ref{eq:T-Q-1}) the determinant $W_{-1,1,\cdots,r}^{(r+1)}$ is the T-function. Here we present the relation to the T-function more precisely.
Choosing the basis of the solutions to the linear problem in $V^{(1)}$ as $\{s^{(1)}_{-r+1},s^{(1)}_{-r+2},\cdots, s^{(1)}_j,\cdots, s^{(1)}_0,s^{(1)}_1\}$, we expand $s_k^{(1)}$ as
\beeq\label{eq:sk-expansion}
s^{(1)}_{k}=(-1)^{r}{\cal T}_{1,k-2}^{[k]}s^{(1)}_{-r+1}
+\sum_{j=1}^r(-1)^{j-1}{\cal T}_{j,k-1}^{[k-1]}s^{(1)}_{-j+2}\\
\eeq
where 
\bee\label{eq:Cal-T-A_r}
{\cal T}_{1,m}&=&\langle s^{(1)}_{-r+1},s^{(1)}_{-r+2},\cdots, s^{(1)}_{0},s^{(1)}_{m+1}\rangle^{[-m]}\\
{\cal T}_{j,m}&=&\langle s^{(1)}_{-r+1},s^{(1)}_{-r+2},\cdots,s^{(1)}_{-j+1},s^{(1)}_{-j+3},\cdots, s^{(1)}_{1},s^{(1)}_{m+1}\rangle^{[-m]},\quad j=2,\cdots,r.
\ee
Here the superscript ${[k]}$ means the shift of spectral parameter $\lambda$, $f^{[\ell]}(\lambda):=f(\lambda+\frac{\ell}{2}\frac{ 2\pi i}{hM})$. 
We find that  $W_{-1,1,\cdots,r}^{(r+1)}$ in (\ref{eq:T-Q-1}) can be written as
\bee
W_{-1,1,\cdots,r}^{(r+1)}{\det}({\cal X}^{(1)}_1,\cdots,{\cal X}^{(1)}_{r+1})=(-1)^{r}{\cal T}_{1,-r-2}^{[r-2]}.
\ee

For the $A_1^{(1)}$ case with $g=0$, $T_{m}(\lambda):={\cal T}_{1,m}^{[-1]}$ satisfy the $(A_1, A_{2M-1})$-type T-system.
For $g\neq 0$, the T-system changes from the $(A_1, A_{2M-1})$-type due to the monodromy of the solutions around the origin
\cite{Lukyanov:2010rn}.  Here we have considered the case of $M$ being integer or half-integer.
For generic non-rational $M$, the T-system becomes semi-infinite $(A_1,A_{\infty})$. 

For the $A_r^{(1)}$ $(r\geq 2)$ case with $g=0$, we can define the T-functions from ${\cal T}_{a,m}$ ($a=1,\cdots, r$).
For generic non-rational $M$, the T-system is of the type $(A_r,A_{\infty})$. For the case with $hM$ being integer, the T-system truncates to the $(A_r,A_{hM-1})$-type. For the $g\neq 0$ case, the T-system becomes more complicated due to the monodromy, where we have studied a similar problem for the $B_2^{(1)}$ case \cite{Ito:2016qzt}.
We note that for $A_2^{(1)}$ it includes  the constant solution of \cite{Saleur:2000bq}.

Choosing other pairs of $(i_0,i_1,\cdots,i_{p-1},i_p)$ in the Pl\"ucker relations,  we can obtain
various T-Q relations.
Taking the conformal limit, which will be discussed in the next section, we can obtain the T-Q
relations which were found in \cite{Bazhanov:2001xm} for the $W_3$ algebra. Substituting (\ref{eq:small-solution}) and (\ref{eq:Q-function}) into (\ref{eq:sk-expansion}), and taking the conformal limit, we obtain the Baxter T-Q relations in \cite{Kojima:2008zza} for the $W_{r+1}$ algebra.
\if0
For $A_2$ case, the conformal version of our T-Q relations coincide with (5.10) and (5.11) in \cite{Bazhanov:2001xm}, which is obtained from the $W_3$ algebra. 
Substituting (\ref{eq:small-solution}) and (\ref{eq:Q-function}) into (\ref{eq:sk-expansion}), we obtain 
\bee
0&=&Q_{i}^{[h]}+(-1)^{h}Q_{i}^{[-h]}+(-1)^{h-1}{\cal T}_{r,1}^{[r-2]}Q_{i}^{[2-h]}+\cdots+(-1)^{j}{\cal T}_{j,1}^{[r-2]}Q_{i}^{[-2j+h]}\\
&&
\quad\qquad\qquad\qquad\qquad\qquad\qquad\qquad+\cdots+{\cal T}_{2,1}^{[r-2]}Q_{i}^{[h-4]}-{\cal T}_{1,1}^{[r-2]}Q_{i}^{[2r-h]},\no
\ee
The conformal version of this equation can be found in \cite{Kojima:2008zza},
which is obtain from the $W_h$ algebra
\fi

%%%%%%%%%%%%%%%%%%%%%%%%%%%%%%%%%%%%%%%%%%%%%%%%%%%%%%%%%%%%%%%%%%%%%%%%

\section{Light-cone limit of Bethe ansatz equations}\label{sec:Reduction-Q}
To clarify the analytical properties of $Q^{(a)}(\lambda)$, we study the light-cone limit of the linear problem. 
In this limit the massive ODE/IM correspondence  reduces to the ``massless'' ODE/IM correspondence.
{In the representation $V^{(1)}$,} we express 
 the solution of the first order holomorphic equation of the linear problem (\ref{eq:LP-2})  {in terms of the top component of $\Psi$ } as 
\beeq\label{eq:psi-z}
\Psi=\left(\begin{array}{c}
e^{\frac{\beta}{2}h_{1}^{(1)}\cdot\phi}\tilde{\psi}_{1}\\
-\frac{1}{me^{\lambda}}e^{\frac{\beta}{2}h_{2}^{(1)}\cdot\phi}D(h_{1}^{(1)})\tilde{\psi}_{1}\\
\vdots\\
(-\frac{1}{me^{\lambda}})^{r}e^{\frac{\beta}{2}h_{r+1}^{(1)}\cdot\phi}D(h_{r}^{(1)})\cdots D(h_{1}^{(1)})\tilde{\psi}_{1}
\end{array}\right),
\eeq
where $D(a):=\partial_z+\beta a\cdot \partial_z\phi$ {\cite{Ito:2013aea}}.
The linear problem then reduces to the $(r+1)$-th order differential equation for $\tilde{\psi}_1$
\beeq\label{eq:ODE-z-Ar}
D(h_{r+1}^{(1)})\cdots D(h_{1}^{(1)})\tilde{\psi}_{1}=(-me^{\lambda})^{h}p(z)\tilde{\psi}_1.
\eeq
We now consider the holomorphic light-cone limit. We first take the limit $\bar{z}\to 0$, and
then $z\sim s\to 0, \lambda\to \infty$ with fixed
\beeq
y=(me^{\lambda})^{\frac{1}{M+1}}z,\quad E=s^{hM}(me^{\lambda})^{\frac{hM}{M+1}}.
\label{eq:lcvariables1}
\eeq
In this limit,  (\ref{eq:ODE-z-Ar}) reduces to 
\beeq\label{eq:ODE-red-Ar}
(-1)^{h+1}\left(\partial_{y}+\frac{\beta h_{r+1}^{(1)}\cdot g}{y}\right)\cdots\left(\partial_{y}+\frac{\beta h_{2}^{(1)}\cdot g}{y}\right)\left(\partial_{y}+\frac{\beta h_{1}^{(1)}\cdot g}{y}\right)\tilde{\psi}_1(y)+(y^{hM}-E)\tilde{\psi}_1(y)=0.
\eeq
From a solution $\tilde{\psi}_1(y)$ of (\ref{eq:ODE-z-Ar}), one finds the solution $\Psi$ of the holomorphic 
linear problem  by (\ref{eq:psi-z}).
As $y\to \infty$, the fastest decaying solution along the positive part of the real axis behaves as
\beeq\label{eq:psi-1-Asy}
\tilde{\psi}_1(y)\sim y^{-rM/2}\exp\left(-\frac{y^{M+1}}{M+1}\right)\quad (y\to\infty),
\eeq 
for $M>\frac{1}{h-1}$.
As $y\to 0$, one obtains a set of the basis $\chi_i^{(1)}(y)=y^{-\beta h_i^{(1)}\cdot g+i-1}+\cdots$ ($i=1,2,\cdots,r+1$), where we have chosen ${\chi}_i^{(1)}$ such that it corresponds to ${\cal X}_i^{(1)}$ 
in the holomorphic light-cone limit. In fact we can choose $\tilde{\psi}_1$ such that $\tilde{\psi}_1=%{\sum_{j=1}^{r+1}} 
a_i(g,m)e^{\frac{M\lambda}{M+1}(-\beta h_{i}^{(1)}\cdot g+i-1)}{\chi}_{i}^{(1)}$ gives $\Psi={\cal X}_i^{(1)}$ for some nozero coefficients $a_i(g,m)$. We then expand the solution (\ref{eq:psi-1-Asy}) in terms of the basis $\chi_i^{(1)}(y)$
\beeq\label{eq:z-basisi-D}
\tilde{\psi}_1(y,E)=\sum_{i=1}^{r+1}D_i(E){\chi}_i^{(1)}(y).
\eeq
Note that $D_i(E)$ is a function of $E=s^{hM}(me^{\lambda})^{\frac{hM}{M+1}}$. Since  the solution (\ref{eq:z-basisi-D}) corresponds to $s^{(1)}_0$ in (\ref{eq:Q-function}), we find
\begin{align}\label{eq:Qi-Di}
Q_{i}^{(1)}(\lambda)=\frac{1}{a_{i}(g,m)e^{\frac{M\lambda}{M+1}(-\beta h_{i}^{(1)}\cdot g+i-1)}} D_{i}(E).
\end{align}
Then $Q_{i[-j]}^{(1)}(\lambda)$ is represented as
\begin{align}
Q^{(1)}_{i[-j]}(\lambda)=\frac{\omega^{-{i r\over2}}}{a_i(g,m) e^{{M\lambda\over M+1}(-\beta h_i^{(1)}\cdot g+i-1) }} D_{i[-j]}(E),
\label{eq:qfunctsym}
\end{align}
where $\omega=e^{\frac{2\pi i}{h(M+1)}}$ and
\begin{align}
D_{i[-j]}(E)&{\equiv}\omega^{j (-\beta h_i^{(1)}\cdot g+i-1) +{i r\over2}}D_i(\omega^{jh}E).
\end{align}
Substituting (\ref{eq:Qi-Di}) into (\ref{eq:Wronskian}), 
$W^{(p)}_{i_1,i_2,\dots, i_p}(\lambda)$ is proportional to the determinant of the $a\times a$ matrix, whose $(i, \ell)$-element is $D_{i[-i_\ell]}(E)$
\beeq
D^{(a)}_{[i_1,i_2,\dots, i_a]}(E){\equiv}{\rm det}D_{i[-i_\ell]}(E),
\eeq
which has been introduced
in \cite{Dorey:2000ma} from the Wronskians of the solutions to the ODE. 
\if0
\begin{align}
W^{(a)}[\psi_1,]&=\sum_{0\leq j_1<j_2<\dots j_a\leq n}D^{(a)}_{[j_1,j_2,\dots, j_a]}(E)W^{(a)}[\chi_{j_1},\chi_{j_2},\dots, \chi_{j_a}],
\end{align}
where $W^{(a)}[f_1,\dots, f_a]$ is the Wronskian defined by
\begin{align}
W^{(a)}[f_1,\dots, f_a]&={\rm det}
\left(
\begin{array}{cccc}
f_1 & f_2 & \cdots & f_a \\
f'_1 & f'_2 & \cdots & f'_a \\
\vdots & \vdots & & \vdots\\
f^{(a-1)}_1 & f^{(a-1)}_2 & \cdots & f^{(a-1)}_a \\
\end{array}
\right).
\end{align}
\fi
In particular, 
we find that in the holomorphic light-cone limit
\begin{align}
{W}^{(a)}_{0,1,\cdots a-1}(\lambda)
=\left(\frac{\prod_{i=1}^{a}a_{i}^{-1}\omega^{-\frac{ir}{2}}}{e^{\frac{M\lambda}{M+1}[\beta_{a}+a\frac{r}{2}]}}\right)D^{(a)}_{[0,1,\cdots,a-1]}(E) \label{eq:WD-relation}
\end{align}
where $\beta_{a}=\sum_{j=0}^{a-1}(-\beta h_{j+1}^{(1)}\cdot g+j)-a\frac{r}{2}$.
Note that the pre-factor of the r.h.s. of (\ref{eq:WD-relation}) does not effect the zeros of the both sides in (\ref{eq:WD-relation}).  
For a zero $\Lambda_j^{(a)}$ $(j\geq 0$) of $W^{(a)}_{0,1,\cdots,a-1}(\lambda)$, we introduce the corresponding zero of $D^{(a)}_{[0,1,\cdots ,a-1]}(E)$ as $F_{j}^{(a)}=s^{hM}(me^{{\Lambda}_{j}^{(a)}})^{\frac{hM}{M+1}}$.
Since the zeros of the Wronskian are classified by those corresponding to the two light-cone limits, we label $j\geq 0$ for the zeros in the holomorphic light-cone limit and $j\leq -1$ for those in the anti-holomorphic light-cone limit. 

We define $A^{(a)}(\omega^{h\frac{a-1}{2}}E){\equiv}D^{(a)}_{[0,1,\cdots ,a-1]}(E)$. Then (\ref{eq:Wron-BAEs}) leads to the Bethe ansatz equations in the holomorphic light-cone limit
\beeq\label{eq:BAEs-cfl-z}
\prod_{b=1}^{r}\omega^{C_{ab}\beta_{b}}\frac{A^{(b)}(\omega^{-\frac{h}{2}C_{ab}}{E}_{j}^{(a)})}{A^{(b)}(\omega^{\frac{h}{2}C_{ab}}{E}_{j}^{(a)})}=-1,
\eeq
where 
$${E}_{j}^{(a)}=\omega^{h\frac{a-1}{2}}F_{j}^{(a)}$$ 
are the zeros of $A^{(a)}(E)$. 
The Bethe ansatz equations (\ref{eq:BAEs-cfl-z}) can be solved by using the massless  NLIEs \cite{Dorey:2000ma}, from which one  obtains the asymptotic value of the zeros $E_j^{(a)}$.

We next consider {the solution of} the anti-holomorphic part of the linear problem, which is expressed in terms of the bottom component of $\Psi$ as \cite{Ito:2013aea}
\beeq\label{eq:psi-zbar}
\Psi=\left(\begin{array}{c}
(-\frac{1}{me^{\lambda}})^{r}e^{\frac{\beta}{2}h_{1}^{(1)}\cdot\phi}\bar{D}(-h_{1}^{(1)})\cdots\bar{D}(-h_{r+1}^{(1)})\tilde{\bar{\psi}}_{r+1}\\
\vdots\\
-\frac{1}{me^{\lambda}}e^{\frac{\beta}{2}h_{r}^{(1)}\cdot\phi}\bar{D}(-h_{r+1}^{(1)})\tilde{\bar{\psi}}_{r+1}\\
e^{\frac{\beta}{2}h_{r+1}^{(1)}\cdot\phi}\tilde{\bar{\psi}}_{r+1}
\end{array}\right).
\eeq
Then the linear problem reduces to the $(r+1)$-th order ODE with respect to $\bar{z}$: 
\beeq\label{eq:ODE-zbar-Ar}
\bar{D}(-h_{1}^{(1)})\cdots\bar{D}(-h_{r+1}^{(1)})\tilde{\bar{\psi}}_{r+1}=({-}me^{-\lambda})^{h}\bar{p}(\bar{z})\tilde{\bar{\psi}}_{r+1}
\eeq
with $\bar{D}(a)=\partial_{\bar{z}}+\beta a\cdot \partial_{\bar{z}}\phi$.
We consider the anti-holomorphic light-cone limit. We first take $z\to 0$, and then we consider the limit $\bar{z}\sim s\to 0,  \lambda\to -\infty$ with keeping
\beeq
\tilde{y}=(me^{-\lambda})^{\frac{1}{M+1}}\bar{z},\quad\tilde{E}=s^{hM}(me^{-\lambda})^{\frac{hM}{M+1}}
\label{eq:lcvariables2}
\eeq
fixed. 
In this limit, (\ref{eq:ODE-zbar-Ar}) becomes
\beeq\label{eq:ODE-red-Ar-zbar}
(-1)^{h+1}\left(\partial_{\tilde{y}}-\frac{\beta h_{1}^{(1)}\cdot g}{\tilde{y}}\right)\left(\partial_{\tilde{y}}-\frac{\beta h_{2}^{(1)}\cdot g}{\tilde{y}}\right)\cdots\left(\partial_{\tilde{y}}-\frac{\beta h_{r+1}^{(1)}\cdot g}{\tilde{y}}\right)\tilde{\bar{\psi}}_{r+1}(\tilde{y})+(\tilde{y}^{hM}-\tilde{E})\tilde{\bar{\psi}}_{r+1}(\tilde{y})=0.
\eeq
From a solution $\tilde{\bar{\psi}}_{r+1}(\tilde{y})$ of (\ref{eq:ODE-red-Ar-zbar}), one finds the solution $\Psi$ of the anti-holomorphic 
linear problem  by (\ref{eq:psi-zbar}).
As $\tilde{y}\to \infty$, the fastest decaying solution along the positive part of the real axis behaves as
\beeq\label{eq:Asy-psi-r+1}
\tilde{\bar{\psi}}_{r+1}(y,E)\sim \tilde{y}^{-rM/2}\exp\left(-\frac{\tilde{y}^{M+1}}{M+1}\right) \quad (\tilde{y}\to\infty).
\eeq
As $\tilde{y}\to 0$, one obtains a set of basis $\tilde{\chi}^{(1)}_{i}=\tilde{y}^{\beta h_{i}^{(1)}\cdot g+h-i}+\cdots$ ($i=1, 2,\cdots, r+1$). $\tilde{\bar{\psi}}_{r+1}=\tilde{a}_{i}(g,m)e^{-\lambda\frac{M}{M+1}(\beta h_{i}^{(1)}\cdot g+h-i)}\tilde{\chi}_{i}^{(1)}$ corresponds to $\Psi={\cal X}_i^{(1)}$ for some coefficients $\tilde{a}_i(g,m)$. We then expand the solution (\ref{eq:Asy-psi-r+1}) in terms of the basis $\tilde{\chi}_i^{(1)}(\tilde{y})$.
\beeq\label{eq:barz-basisi-D}
\tilde{\bar{\psi}}_{r+1}(\tilde{y},\tilde{E})=\sum_{i=1}^h\tilde{D}_i(\tilde{E})\tilde{\chi}_i^{(1)}(\tilde{y}).
\eeq
Since (\ref{eq:barz-basisi-D}) corresponds to $s^{(1)}_0$ in (\ref{eq:Q-function}), we find
\begin{align}\label{eq:Qi-tildeDi}
Q_{i}^{(1)}(\lambda)=\frac{1}{\tilde{a}_{i}(g,m)e^{-\frac{M\lambda}{M+1}(\beta h_{i}^{(1)}\cdot g+h-i)}}\tilde{D}_{i}(\tilde{E}).
\end{align}
Then $Q_{i[-j]}^{(1)}(\lambda)$ is represented as 
\begin{align}
Q_{i[-j]}^{(1)}(\lambda)=\frac{\omega^{\frac{ir}{2}}}{\tilde{a}_{i}(g,m)e^{-\frac{M\lambda}{M+1}(\beta h_{i}^{(1)}\cdot g+h-i)}}\tilde{D}_{i[-j]}(\tilde{E}),
\end{align}
where 
\begin{align}
\tilde{D}_{i[-j]}(\tilde{E})&{\equiv}\omega^{-j(\beta h_{i}^{(1)}\cdot g+h-i)-\frac{ir}{2}}\tilde{D}_{i}(\omega^{-jh}\tilde{E}).
\end{align}
Substituting (\ref{eq:Qi-tildeDi}) into (\ref{eq:Wronskian}), $W^{(p)}_{i_1,i_2,\cdots,i_p}(\lambda)$ is proportional to the determinant 
\begin{align}
\tilde{D}^{(a)}_{[0,1,\cdots,a-1]}(\tilde{E}){\equiv}\det \tilde{D}_{i[-i_\ell]}(\tilde{E}).
\end{align}
We find that in the anti-holomorphic light-cone limit
\begin{align}
W_{0,1,\cdots,a-1}^{(a)}(\lambda)=\left(\frac{\prod_{i=1}^{a}\tilde{a}_{i}^{-1}\omega^{\frac{ir}{2}}}{e^{-\lambda\frac{M}{M+1}[\tilde{\beta}_{a}+a\frac{r}{2}]}}\right)\tilde{D}^{(a)}_{[01,\cdots,a-1]}(\tilde{E}),  \label{eq:W-D-barz}
\end{align}
where $\tilde{\beta}_{a}=\sum_{j=0}^{a-1}(\beta h_{j+1}^{(1)}\cdot g+h-j-1)-a\frac{r}{2}$.
Note that the pre-factor of the r.h.s. of (\ref{eq:W-D-barz}) does not effect the zeros of the two sides in (\ref{eq:W-D-barz}). For a zero $\Lambda_j^{(a)} (j\leq -1)$ of $W^{(a)}_{0,1,\cdots,a-1}(\lambda)$, we introduce the zero of $\tilde{D}^{(a)}_{[0,1,\cdots,a-1]}(\tilde{E})$ as $\tilde{F}_j^{(a)}=s^{hM}(me^{-{\Lambda}_j^{(a)}})^{\frac{hM}{M+1}}$.

We introduce $\tilde{A}^{(a)}(\omega^{-h\frac{a-1}{2}}\tilde{E}){\equiv}\tilde{D}_{[0,1,\cdots,a-1]}^{(a)}(\tilde{E})$. Then (\ref{eq:Wron-BAEs}) thus lead to the Bethe ansatz equations for anti-holomorphic light-cone limit
\bee\label{eq:BAEs-cfl-barz}
\prod_{b=1}^{r}\omega^{C_{ab}\tilde{\beta}_{b}}\frac{\tilde{A}^{(b)}(\omega^{-\frac{h}{2}C_{ab}}\tilde{E}_{j}^{(a)})}{\tilde{A}^{(b)}(\omega^{\frac{h}{2}C_{ab}}\tilde{E}_{j}^{(a)})}=-1,
\ee 
where $$\tilde{E}_{j}^{(a)}
=\omega^{-h\frac{a-1}{2}}\tilde{F}_{j}^{(a)}%=s^{hM}(me^{-(\tilde{\Lambda}_{j}^{(a)}+\frac{a-1}{2}\frac{2\pi i}{hM})})^{\frac{hM}{M+1}},
$$ are the zeros of $\tilde{A}^{(a)}(\tilde{E})$.
Note that the Bethe ansatz equations (\ref{eq:BAEs-cfl-barz}) and the zeros $\tilde{E}_j^{(a)}$ can be obtained from the the Bethe ansatz equations (\ref{eq:BAEs-cfl-z}) and the zeros $E_j^{(a)}$ respectively by replacing $E\to \tilde{E}$ and $\beta_m\to \tilde{\beta}_m$.

We can also introduce ${D}^{(a)}_{[0,1,\cdots,a-1]}({E})$  and $\tilde{D}^{(a)}_{[0,1,\cdots,a-1]}(\tilde{E})$ from the Wronskian of $\tilde{\psi}_1(y)$ and $\tilde{\bar{\psi}}_{r+1}(\tilde{y})$ respectively.
Using the Pl\"ucker relation of the Wronskians, the Bethe Ansatz equations (\ref{eq:BAEs-cfl-z}) and (\ref{eq:BAEs-cfl-barz}) are obtained.

In summary, the zeros of the Q-functions come from those the $D$-function obtained in the two light-cone limits of the linear problems.  In the next section we will determine the analytical structure of the Q-functions using the structure of zeros.

%%%%%%%%%%%%%%%%%%%%%%%%%%%%%%%%%%%%%%%%%%%%%%%%%%%%%%%%%%%%%%%%%%%%

\section{Non-linear integral equations}\label{sec:NLIEs-Ar}

In this section we derive the non-linear integral equations from the Bethe ansatz equations (\ref{eq:BAE-psi}). 
First we discuss analytic properties of the Q-function, which are determined by their asymptotic properties and zeros. Then we introduce the counting functions,
and derive the non-linear integral equations satisfied by the counting functions. These
equations provide a basic tool to investigate the massive ODE/IM correspondence.

\subsection{Asymptotic behavior of $Q_1^{(a)}(\lambda)$ at large $|\lambda|$}
We study the asymptotic behavior of $Q_1^{(a)}(\lambda)$ at large $|\lambda|$.
Since $Q_1^{(a)}(\lambda)$ do not depend on the coordinates $z,\bar{z}$, one can use the solutions of the linear problem around $z=0$ 
to evaluate them.
In particular, $Q^{(1)}_1(\lambda)$ is evaluated as
\begin{eqnarray}\label{eq:Q_1-A_r}
	Q^{(1)}_1(\lambda)=\braket{s^{(1)}_{0},{\cal X}^{(1)}_{2},\cdots,{\cal X}^{(1)}_{r+1}}=\lim_{|z|\to 0}\psi_{1}e^{\beta(\lambda+i\theta)g\cdot h^{(1)}_{1}},
\end{eqnarray}
where $\psi_1$ is the first component of $s_0^{(1)}$.  

The asymptotic behavior of  the solution $\tilde{\psi}_1$ to  (\ref{eq:ODE-z-Ar}) at ${\rm Re}(\lambda)\to \infty$ can be obtained by the WKB method. The result is 
\begin{align}\label{eq:Asy-psi1}
\tilde{\psi}_{1}&\sim\exp\left[me^{\lambda}\int_{\log z}^{\infty}dx'\left(e^{x^{\prime}}(e^{hMx^{\prime}}-s^{hM})^{\frac{1}{h}}-e^{(M+1)x^{\prime}}\right)\right],\quad {\rm Re}(\lambda)\to \infty.
\end{align}
Then substituting (\ref{eq:Asy-psi1}) to (\ref{eq:Q_1-A_r}), and taking the limit $\log z\to -\infty$, we obtain
\beeq\label{eq:Asy-Q-lambda>0}
\log Q^{(1)}_1(\lambda)\to (-E)^{\frac{M+1}{hM}}\kappa(hM,h)\qquad \mbox{for}\quad {\rm Re}(\lambda)\to \infty,\quad |\mbox{arg}(-E)|<\pi,
\eeq
where $E$ is defined in (\ref{eq:lcvariables1}). $\kappa(a,b)$ is
\bee
\kappa(a,b)=\int_{0}^{\infty}dx[(x^{a}+1)^{1/b}-x^{a/b}]=\frac{\Gamma(1+\frac{1}{a})\Gamma(1+\frac{1}{b})\sin(\frac{\pi}{b})}{\Gamma(1+\frac{1}{a}+\frac{1}{b})\sin(\frac{\pi}{a}+\frac{\pi}{b})}.
\ee

For ${\rm Re}(\lambda)\to -\infty$, from the WKB solution $\tilde{\bar{\psi}}_{r+1}$ to the anti-holomorphic ODE (\ref{eq:ODE-zbar-Ar}), we calculate the behavior of the first component of (\ref{eq:psi-zbar}). 
The asymptotic behavior $Q^{(1)}(\lambda)$ at ${\rm Re}(\lambda)\to -\infty$ is then evaluated by using (\ref{eq:Q_1-A_r}) as
\beeq\label{eq:Asy-Q-lambda<0}
\log Q^{(1)}_1(\lambda)\to (-\tilde{E})^{\frac{(M+1)}{hM}}\kappa(hM,h),\qquad{\rm Re}(\lambda)\to-\infty,\quad|\mbox{arg}(-\tilde{E})|<\pi
\eeq
where $\tilde{E}$ is defined in (\ref{eq:lcvariables2})\footnote{We have assumed that $s$ is  real. }.
$Q^{(1)}_i(\lambda)$ at ${\rm Re}(\lambda)\to \infty$ and ${\rm Re}(\lambda)\to -\infty$ are the same as (\ref{eq:Asy-Q-lambda>0}) and (\ref{eq:Asy-Q-lambda<0}) respectively at the leading order. 

Using the $\psi$-system (\ref{eq:first-component-psi-system}), the Wronskian of $Q_1^{(a)}(\lambda)$ (\ref{eq:Q1a-Qa})  and (\ref{eq:Q2a-Qa}), we determine the asymptotics of $Q^{(a)}_1(\lambda)$ as 
\bee
\log Q^{(a)}_1(\lambda)&\to&\frac{\sin(\frac{a\pi}{{h}})}{\sin(\frac{\pi}{{h}})}(-E)^{\frac{(M+1)}{hM}}\kappa(hM,h),\quad{\rm Re}(\lambda)\to\infty,\quad |\mbox{arg}(-E)|<\pi,  \label{eq:LAsy-Qm-Ar}\\
\log Q^{(a)}_1(\lambda)&\to&\frac{\sin(\frac{a\pi}{{h}})}{\sin(\frac{\pi}{{h}})}(-\tilde{E})^{\frac{(M+1)}{hM}}\kappa(hM,h),\quad {\rm Re}(\lambda)\to-\infty,\quad |\mbox{arg}(-\tilde{E})|<\pi.\label{eq:-LAsy-Qm-Ar}
\ee

\subsection{Zeros of $Q^{(a)}(\lambda)$}
In the previous section, we have studied the zeros 
of $Q_1^{(a)}(\lambda)$. We considered the holomorphic light-cone limit, in which case $Q_1^{(a)}(\lambda)$ reduces to $A^{(a)}(E)$, whose zeros are ${E}_{j}^{(a)}=s^{hM}(me^{\hat{\lambda}_{j}^{(a)}})$. 
As observed in the previous section, $Q^{(a)}_1(\lambda)$ and the $A^{(a)}(E)$ have the same asymptotic value of zeros, i.e. $\lambda^{(a)}_j\to \hat{\lambda}^{(a)}_j$ in the holomorphic light-cone limit. 
For large $E$, the asymptotic value of ${E}^{(a)}_{j}$ tends to ${\cal E}_{j}^{(a)}$, which is defined by \cite{Dorey:2000ma}:
\beeq
{E}^{(a)}_{j}\to {\cal E}_{j}^{(a)} 
{\equiv}\left\{\frac{\sin(\frac{\pi}{h})}{\sin(\frac{\pi a}{h})}\frac{\pi}{b_{0}M_a}[2j+1+\hat{\alpha}_{a}({g})]\right\}^{\frac{hM}{M+1}},\quad j\to \infty,\label{eq:zeros-asymptotic}
\eeq
where
\beeq\label{eq:Asy-b-M}
b_{0}=2\sin\left(\pi\frac{(M+1)}{hM}\right)\kappa(hM,h),\quad M_{a}=ms^{M+1}\frac{\sin(\frac{a\pi}{h})}{\sin(\frac{\pi}{h})}.
\eeq
The parameter $\hat{\alpha}_a({g})$ is defined as
\beeq
\hat{\alpha}_{a}({g})=-\frac{2}{h}\beta_a.
\eeq
We consider the anti-holomorphic light-cone limit, in which case $Q_1^{(a)}(\lambda)$ reduces to $\tilde{A}^{(a)}(\tilde{E})$, whose zeros are labeled by $\tilde{E}_{j}^{(a)}=s^{hM}(me^{-\tilde{\lambda}_{j}^{(a)}})$. $Q^{(a)}_1(\lambda)$ and the $\tilde{A}^{(a)}(\tilde{E})$ have the same asymptotic values of zeros, i.e. $\lambda^{(a)}_j\to \tilde{\lambda}^{(a)}_j$ in this limit.  
At large $\tilde{E}$, $\tilde{E}_{j}^{(a)}$ tends to $\tilde{\cal E}_{j}^{(a)}$ \cite{Dorey:2000ma}, where
\bee
\tilde{E}_{j}^{(a)}\to \tilde{\cal{E}}_{j}^{(a)}
=\left\{\frac{\sin(\frac{\pi}{h})}{\sin(\frac{\pi m}{h})}\frac{\pi}{b_{0}}[2(-j-1)+1-\hat{\alpha}_{a}({g})]\right\}^{\frac{hM}{M+1}},\quad j\to -\infty \label{eq:zeros-asymptotic-1}.
\ee

Thus we can read off the zeros of $Q^{(a)}_1(\lambda)$ by using the ones of $A^{(a)}(E)$ or $\tilde{A}^{(a)}(\tilde{E})$, which is obtained in the light-cone limit.
This limit means the parameter $s$ should be small, where $\lambda^{(a)}_j\to \hat{\lambda}^{(a)}_j (\tilde{\lambda}^{(a)}_j)$ in the light-cone limit. 
As the value of zeros $\lambda_j^{(a)}$ changes with increasing $s$, no additional zeros can be generated because the asymptotic formulas (\ref{eq:LAsy-Qm-Ar}) and (\ref{eq:-LAsy-Qm-Ar}) are valid for any $s$.
%It is convenient to label the zeros $\tilde{E}_j^{(a)}$ as follows: 
%\beeq
%e^{\lambda_{j}^{(a)}\frac{hM}{M+1}}=\begin{cases}
%m^{-\frac{hM}{M+1}}s^{-hM}E_{j}^{(a)}[\hat{\alpha}_{a}({g})], & j\geq0\\
%m^{\frac{hM}{M+1}}s^{hM}\frac{1}{\tilde{E}_{-j-1}^{(a)}[-\hat{\alpha}_{a}({g})]}, & j<0
%\end{cases},
%\eeq
%where we have summarized the zeros as a set $\{E_{j}^{(a)}(\pm \hat{\alpha}_a)\}|_{j=0}^{\infty}$.

For $M>\frac{1}{h-1}$, the order $\frac{M+1}{hM}$ of the functions $Q^{(a)}(\lambda)$ in (\ref{eq:LAsy-Qm-Ar}) and (\ref{eq:-LAsy-Qm-Ar}) is less than one. The Hadamard factorization theorem %of $Q$-function 
leads to
\beeq
Q_1^{(a)}(\lambda)=G^{(a)}({g})e^{\frac{hM\lambda}{2(M+1)}\hat{\alpha}_{a} }\prod_{j=0}^{\infty}(1-e^{\frac{hM}{M+1}(\lambda-\lambda_{j}^{(a)})})\prod_{j=-\infty}^{-1}(1-e^{-\frac{hM}{M+1}(\lambda-\lambda_{j}^{(a)})}),\label{eq:Hadamard-factorization}
\eeq
where $G^{(a)}({g})$ is a constant.

\subsection{Non-linear Integral Equations}
Let us introduce the counting function $a^{(a)}(\lambda)$%\footnote{This function reduce to $a^{(m)}$ in \cite{Dorey:2000ma}.}
\beeq\label{eq:counting-fun}
a^{(a)}(\lambda)=\prod_{b=1}^{r}%e^{-C_{mt}\frac{2\pi i}{hM}\beta_{t}}
\frac{Q_1^{(b)}(\lambda+\frac{\pi i}{hM}C_{ab})}{Q_1^{(b)}(\lambda-\frac{\pi i}{hM}C_{ab})},\quad a=1,2,\cdots,r.
\eeq
The function satisfies $a^{(a)}(\lambda_j^{(a)})=-1$ for zeros $\lambda^{(a)}_j$ of $Q_1^{(a)}(\lambda)$.
Then we use (\ref{eq:Hadamard-factorization}) and the procedure in \cite{Destri:1992ey,Destri:1994bv} to rewrite (\ref{eq:counting-fun}) as
\beeq
\log a^{(a)}(\lambda)=\sum_{b=1}^{r}\frac{\pi i}{(M+1)}C_{ab}\hat{\alpha}_{b}({g})+\sum_{b=1}^{r}\int_{{\cal C}}d\lambda'F_{ab}(\lambda-\lambda')\partial_{\lambda'}\log(1+a^{(b)}(\lambda')),
\eeq
where the integral contour ${\cal C}$ encircles all the zeros anti-clockwise, and the kernel $F_{ab}(\lambda)$ is defined by
\beeq\label{eq:Fmt}
F_{ab}(\lambda)=\log\left[\frac{\sinh[\frac{1}{2}\frac{hM}{M+1}\lambda-\frac{1}{2}\frac{\pi iM}{M+1}C_{ab}]}{\sinh[\frac{1}{2}\frac{hM}{M+1}\lambda+\frac{1}{2}\frac{\pi iM}{M+1}C_{ab}]}\right].
\eeq
Here we assume all the zeros of $Q_1^{(a)}(\lambda)$ are real 
as observed in the analysis of the ODE, which corresponds to  the ground state of the
Bethe ansatz equations \cite{Dorey:2000ma}.
\if0
\footnote{In the light-cone limit, the linear problem leads to the ODEs (\ref{eq:ODE-red-Ar}) and (\ref{eq:ODE-red-Ar-zbar}). The zeros $\lambda_j$ of $Q^{(a)}_j(\lambda)$ are corresponding to the zeros $E^{(a)}_j$ or $\tilde{E}^{(a)}_j$. It was assumed in \cite{Dorey:2000ma} that both $E^{(a)}$ and $\tilde{E}^{(a)}$ are positive real under certain boundary condition. Since our boundary for ODE is same as the one in \cite{Dorey:2000ma}, our $\lambda^{(a)}_j$ is real too under this assumption.
As $|z|\to 0$, from linear problem we obtain (\ref{eq:ODE-red-Ar}) and (\ref{eq:ODE-red-Ar-zbar}) as well, which should be considered at the same time. We could also introduce the corresponding $A^{(E)}$ and $\tilde{A}^{(\tilde{E})}$ as in the lightcone limit case. Our Bethe ansatz equations lead to the Bethe ansatz equation about $A(E)$ or $\tilde{A}(\tilde{E})$.
If $\lambda^{(a)}$ are the zeros of $Q^{(1)}_1(\lambda)$, the corresponding $E^{(a)}_j$ or $\tilde{E}^{(a)}_j$ should also be the zeros of the corresponding $A^{(a)}(E)$ or $\tilde{A}^{(a)}(\tilde{E})$. Under the same assumption in \cite{Dorey:2000ma}, our $\lambda^{(a)}_j$ is real.}, which corresponds to the ground state of the Bethe ansatz equation.
\fi
Then the definition of $a^{(a)}(\lambda)$ leads to $(a^{(a)}(\lambda))^{\ast}=(a^{(a)}(\lambda^{\ast}))^{-1}$. 
Then integrating by parts and taking the Fourier transformation 
${\cal F}[f](k)=\int_{-\infty}^{\infty}f(\lambda)e^{-i\lambda k}d\lambda$, we obtain
\beeq
\sum_{b=1}^{r}(\delta_{ab}-{\cal F}[R_{ab}]){\cal F}[\log a^{(b)}
%(\lambda)
]
=\left[\sum_{b=1}^{r}\frac{\pi i}{(M+1)}C_{ab}\hat{\alpha}_{b}({g})\right]2\pi\delta(k)-\sum_{b=1}^{r}2i{\cal F}[R_{ab}]{\cal F}\left[{\rm Im}\log[1+a^{(b)}]\right].
\eeq
where $R_{ab}(\lambda-\lambda')=\frac{i}{2\pi}\partial_{\lambda-\lambda'}F_{ab}(\lambda-\lambda')$.
Applying the inverse matrix of $(\mathbf{1}-{\cal F}[R])$ to this equation and taking the inverse Fourier transformation, we obtain the NLIEs:
\bee\label{eq:DDV-Ar}
\log a^{(a)}(\lambda)&=&-2ib_0M_a\sinh\lambda+i\pi\gamma_{a}+\sum_{b=1}^{r}\int_{{\cal C}_{1}}d\lambda'\varphi_{ab}(\lambda-\lambda')\log[1+a^{(b)}](\lambda')\no\\
&&-\sum_{b=1}^{r}\int_{{\cal C}_{2}}d\lambda'\varphi_{ab}(\lambda-\lambda')\log%\{
\left[1+\frac{1}{a^{(b)}(\lambda')}\right],
\ee
where ${\cal C}_1$ (${\cal C}_2$) runs from $-\infty-i0$ ($\infty+i0$) to $\infty-i0$ ($\infty+i0$) and
\bee
\varphi_{ab}(\lambda)&=&{\cal F}^{-1}\left[{\bf 1}-(1-{\cal F}[R])_{ab}^{-1}\right]=-{\cal F}^{-1}\left[(1-{\cal F}[R])_{ac}^{-1}{\cal F}[R_{cb}]\right]\label{eq:varphi_mt}\\
i\pi{\gamma}_{a}&=&\sum_{c,b}^{r}{\cal F}^{-1}\left[(\delta_{ac}-{\cal F}[R_{ac}])^{-1}\frac{\pi i}{(M+1)}C_{cb}\hat{\alpha}_{b}2\pi\delta(k)\right].
\ee
The driving term $-2ib_0M_a\sinh\lambda$ is due to the zeros modes of $(\mathbf{1}-{\cal F}[R](k))^{-1}$ at $k=i$, $b_0$ and $M_m$ are defined in (\ref{eq:Asy-b-M}).
From (\ref{eq:Fmt}), we obtain the non-vanishing ${\cal F}[R_{ab}](k)$
\bee
{\cal F}[R_{ab}](k)=\frac{\sinh[\frac{\pi k}{hM}((M-1)\delta_{ab}+\delta_{a,b+1}+\delta_{a,b-1})]}{\sinh(\frac{M+1}{hM}k\pi)}.
\ee
Then it is easy to find $\gamma_a=\hat{\alpha}_a$.
To evaluate $\varphi_{ab}(\lambda)$, we introduce the generalized Cartan matrix \cite{ZinnJustin:1997at} 
\bee
C_{ab}(k)&=&2\delta_{ab}-\frac{1}{\cosh[\frac{\pi k}{h}]}(\delta_{a,b-1}+\delta_{a,b+1})\\
C_{ab}^{-1}(k)&=&C_{ba}^{-1}(k)=\frac{\coth(\frac{\pi}{h}k)\sinh[\frac{\pi}{h}(h-a)k]\sinh[\frac{\pi}{h}bk]}{\sinh(\pi k)}\quad(a\geq b).
\ee
Then (\ref{eq:varphi_mt}) can be written as
\bee
\varphi_{ab}(\lambda)=\frac{1}{2\pi}\int_{-\infty}^{\infty}dke^{ik\lambda}\left[\delta_{ab}-\frac{\sinh[\frac{\pi}{h}(1+\xi)k]}{\cosh[\frac{\pi}{h}k]\sinh[\frac{\pi}{h}\xi k]}C_{ab}^{-1}(k)\right].
\ee
In Appendix A, the NLIEs  (\ref{eq:DDV-Ar}) are shown to be equivalent to those in
\cite{ZinnJustin:1997at}.
%
%After identifying some parameters, we find the NLIEs (\ref{eq:DDV-Ar}) exactly match with the ones in \cite{ZinnJustin:1997at}. 
%We will discuss the correspondence in Appendix \ref{sec:massive-IM}.

\subsection{UV limit %of massive integrable model
}
For convenience, we introduce other counting functions $Z_a(\lambda)$ and $\tilde{Q}_{a}(\lambda)$ $(a=1,2,\cdots,r)$ as
\beeq
e^{-iZ_{a}(\lambda)}{\equiv} a^{(a)}(\lambda),\qquad \tilde{Q}_{a}(\lambda){\equiv}\frac{1}{i}\log\frac{1+
e^{iZ_{a}(\lambda+i0)}}{1+e^{-iZ_{a}(\lambda-i0)}}.
\eeq
The NLIEs (\ref{eq:DDV-Ar}) are written in terms of $Z_a(\lambda)$ and $\tilde{Q}_{a}(\lambda)$ as (\ref{eq:NLIE-Zinn-Justin}) in Appendix \ref{sec:massive-IM}.
In the UV limit $2b_0M_1\to 0$, the corresponding massive integrable model flows to its UV fixed point. 
In this limit, the NLIEs split into three types of equations corresponding to  two asymptotic regions and one intermediate region, where two asymptotic regions are separated by the distance $\log\frac{1}{b_0M_1}$. 
In the intermediate region, $Z_{a}(\lambda)$ is flat. 
In the two asymptotic regions which are in both sides,
the two decoupled counting functions
\bee
Z_{a}^{\pm}(\lambda){\equiv}\lim_{2b_0M_1\to0}Z_{a}(\lambda\pm\log\frac{1}{b_0M_1}),\quad \tilde{Q}_{a}^{\pm}(\lambda){\equiv}\frac{1}{i}\log\frac{1+%(-1)^{\delta_{s}}
e^{iZ_{a}^{\pm}(\lambda+i0)}}{1+%(-1)^{\delta_{s}}
e^{-iZ_{a}^{\pm} (\lambda-i0)}}.
\ee
are defined.
Then the NLIEs become
\beeq
Z_{a}^{\pm}(\lambda)=-\pi\hat{\alpha}_{a}\pm\frac{M_{a}}{M_1}e^{\pm\lambda}+\sum_{b=1}^{r}X_{ab}\ast \tilde{Q}_{b}^{\pm}(\lambda).\label{eq:UV-limit-DDV}
\eeq
The asymptotic behaviors of $Z_{a}^{\pm}(\lambda)$ and $\tilde{Q}_a^{\pm}(\lambda)$ for $\lambda=\pm\infty$ are
\beeq
Z_a^\pm(\pm\infty)=\pm \infty,\quad \tilde{Q}_a^{\pm}(\pm \infty)=0.
\eeq
At $\lambda=\mp \infty$, (\ref{eq:UV-limit-DDV}) leads to the constraints on $Z_{a}^{\pm}(\mp\infty)$
\beeq
Z_{a}^{\pm}(\mp\infty)=-\pi\hat{\alpha}_{a}+\sum_{b=1}^{r}[X_{ab}\ast \tilde{Q}_{b}](\mp\infty)
=-\pi\hat{\alpha}_{a}+\sum_{t=1}^{r}\tilde{Q}_{b}(\mp\infty)\chi_{ab}(\infty),
\eeq
where $\chi_{ab}(\infty)=\int_{-\infty}^{\infty}d\lambda X_{ab}(\lambda)=\delta_{ab}-(M+1)C_{ab}^{-1}(k=0)$.
Solving these constraints, we obtain the constant solution of $\tilde{Q}^\pm_a(\mp\infty)$
\beeq\label{eq:cosn-Q-Ar}
\tilde{Q}^\pm_a(\mp\infty)=-\frac{\pi}{M+1} \sum_{b=1}^rC_{ab}\hat{\alpha}_b=-\frac{2\pi}{h(M+1)}(1+\beta g\cdot \alpha_a),
\eeq
where $C_{ab}$ is the Cartan matrix and $\alpha_a$ is the simple root.

The effective central charge is given as
\beeq
c_{eff}(2b_0M_1)=-\frac{6}{\pi^{2}}2b_0\sum_{a=1}^{r}M_{a}\left[\int_{{\cal C}}d\lambda\sinh\lambda{\rm Im}\log(1+a^{(a)}(\lambda))\right].
\eeq
In the UV limit, the effective central charge becomes
\beeq
c_{eff}(0)=\frac{3}{\pi^{2}}\sum_{a=1}^{r}\frac{M_{a}}{M_1}\left[\int_{-\infty}^{\infty}d\lambda\frac{d e^{\lambda}}{d\lambda}\tilde{Q}_{a}^{+}(\lambda)-\int_{-\infty}^{\infty}d\lambda\frac{d (-e^{-\lambda})}{d\lambda}\tilde{Q}_{a}^{-}(\lambda)\right].
\eeq
Using the multi-component generalization \cite{ZinnJustin:1997at} of the lemma in sec.VIII of \cite{Destri:1997yz} 
\if0
\bee
&&\frac{3}{\pi^{2}}\sum_{a=1}^{r}\frac{M_{a}}{M_1}\int d\lambda\frac{d}{d\lambda}(\pm e^{\pm\lambda})\tilde{Q}_{a}^{\pm}(\lambda)\\
&=&\frac{3}{\pi^{2}}\{-2\sum_{a=1}^r{\rm Re}\int_{\Gamma_{m}^\pm}\frac{du}{u}\log(1+u)-\frac{1}{2}\sum_{a,b=1}^{r}[\tilde{Q}_{a}^{\pm}(\mp\infty)\tilde{Q}_{b}^{\pm}(\mp\infty)-\tilde{Q}_{a}^{\pm}(\mp\infty)\tilde{Q}_{b}^{\pm}(\mp\infty)]\chi_{ab}(\infty)
\}\no,
\ee
where $\Gamma_m^\pm$ goes from $Z_m^\pm(-\infty+i0)$ to $Z_m^\pm(+\infty)(+\infty+i0)$ avoiding the logarithmic cut.
The contour $\pm\Gamma_m^\pm=\pm\int_{\tilde{Q}_{m}^{\pm}(-\infty)}^{\tilde{Q}_{m}^{\pm}(-\infty)}$ can be divide as the unit arc from $Z_m^{\pm}(\mp\infty+i0)=\tilde{Q}_m^\pm(\mp\infty)$ to point $1$ and the segment from $1$ to $Z_m^\pm(\pm\infty+i0)=0$, and obtain
\bee
-2{\rm Re}\int_{\Gamma_{m}^{\pm}}\frac{du}{u}\log(1+u)=\pm\left(-(\tilde{Q}_{m}^{\pm})^{2}(\mp\infty)+\frac{\pi^{2}}{6}\right).
\ee

\bee
\frac{3}{\pi^{2}}\sum_{m=1}^{r}\frac{M_{m}}{M_1}\int_{-\infty}^{\infty}d\lambda e^{\pm\lambda}\tilde{Q}_{m}^{\pm}(\lambda)=\frac{3}{\pi^{2}}\left(\pm\sum_{m=1}^{r}\frac{\pi^{2}}{6}\mp\frac{1}{2}\sum_{m,t=1}^{r}\tilde{Q}_{m}^{\pm}(\mp\infty)C_{mt}^{-1}\tilde{Q}_{t}^{\pm}(\mp\infty)(M+1)\right).
\ee
\fi
and the constant $\tilde{Q}^{\pm}_a(\mp\infty)$ in (\ref{eq:cosn-Q-Ar}), we obtain
\bee\label{eq:ceff-1}
c_{eff}(0)&=&r-\frac{3}{M+1}\sum_{a,b=1}^{r}(C\hat{\alpha})_aC^{-1}_{ab}(C\hat{\alpha})_b\no\\
&=&r-\frac{12}{h^2(1+M)}\sum_{a,b=1}^{r}({\bf 1}_a+\beta g\cdot \alpha_a)C_{ab}^{-1}({\bf 1}_b+\beta g\cdot \alpha_b),
\ee
where ${\bf 1}_a=1$. (\ref{eq:ceff-1}) is simplified to
\bee
c_{eff}(0)=r-\frac{12}{h^2(M+1)}(\rho^\vee+\beta g)^2,
\ee
where $\rho^\vee$ is the co-Weyl vector of $A_r$ algebra.
For $g=0$, we find
\bee
c_{eff}^{g=0}(0)=r-\frac{3}{M+1}\frac{r(r+2)}{3(r+1)}=(h-1)\left(1-\frac{(h+1)h}{pq}\right),
\ee
which coincides with the effective central charge of non-unitary CFT $WA_{r}^{(p,q)}$ with $p=r+1=h$ and $q=h(M+1)$ \cite{Fateev:1987zh,Fateev-Lukyanov:review,Dunning:2002cu}.

Let us comment on the Thermodynamic Bethe Ansatz (TBA) equations from the modified affine Toda field equation for $g\neq 0$.
For the simplest case, i.e. the $A_1^{(1)}$-type modified affine Toda field equation, one obtains D-type Y-system \cite{Lukyanov:2010rn} for an integer $2M$. We find the periodic condition  of the $Y-$function from the quasi-periodic condition of $Q_1^{(1)} (\lambda)$.
Especially for the case where $M$ is an integer, the periodic condition and the shift of the spectral parameter of Y-functions coincide with those of $D_{M+1}$-type Y-system \cite{Zamolodchikov:1991et,Ravanini:1992fi}. We can derive the TBA equations from the Y-system and compute the effective central charge in the UV limit. This effective central charge coincides with the one obtained from the NLIE approach.
For a half integer $M$, both the shift of the spectral parameter in the Y-function and the periodic condition do not coincide with the those of the usual D-type Y-system. It is interesting to derive the TBA equation in this case.

%%%%%%%%%%%%%%%%%%%%%%%%%%%%%%%%%%%%%%%%%%%%%%%%%%%%%%%%%%%%%%%%%%%%%%%%%%%%%%%%%%%%%%%%%%%%%%%%%%%%%%%%%%%%%%%%%%%%%%%%%%%%%%%%%%%%%%%%%%%%

%%%%%%%%%%%%%%%%%%%%%%%%%%%%%%%%%%%%%%%%%%%%%%%%%%%%%%%%%%%%%%%%%%%%%%%%%%%%%%%%%%%%%%%%%%%%%%%%%%%%%%%%%%%%%%%%%%%%%%%%%%%%%%%%%%%%%%%%%%%%

\section{Conclusions and discussions}\label{sec:conclusions and discussion}
In this paper, we have studied the massive ODE/IM correspondence between the $A_r^{(1)}$-type modified affine Toda field equations and the two-dimensional massive integrable models.
The Q-functions are introduced from the solutions of the linear problems associated with the modified affine Toda field equations.
The $\psi$-system satisfied by the solutions leads to the Bethe ansatz equations.
The asymptotics of the $Q$-functions for large $\lambda$, the spectral parameter,  is obtained by the WKB solutions and with the help of the $\psi$-system.
We then have derived the Bethe ansatz equations of the massive integrable models.
In the light-cone limit, we found that the correspondence reduces to the relation between the ODE and the
massless integrable model, where
the Q-functions are represented as the Wronskians of the basic Q-functions.
From the Bethe ansatz equations and the asymtotics of the Q-functions, we have derived the non-linear integral equations of $A_r^{(1)}$-type, which agree with the ones obtained in  \cite{ZinnJustin:1997at}.
Based on the NLIEs, we derived the effective central charge in the UV limit, which depends on the monodoromy
parameter $g$ of the solutions of the linear problem around the origin. At $g=0$, the effective central charge at UV limit coincides with the effective central charge of non-unitary CFT $WA_{r}^{(p,q)}$ with $p=r+1=h$ and $q=h(M+1)$. 

In the present paper we have worked out in the case of $A_r^{(1)}$-type affine Toda field equations as a typical example of affine Lie algebras. 
It would be possible to generalize to the affine Lie algebra of $\hat{\mathfrak g}^{\vee}$,  from which we obtain the Bethe ansatz equations associaed with the affine Lie algebra $\hat{\mathfrak g}$ \cite{Ito:2015nla} and the non-unitary $W{\mathfrak g}$-minimal model in the UV-limit.
These will be presented in a separate paper.

These W${\mathfrak g}$-models also appear  in the context of the correspondence between the Argyres-Douglas theories of $(A_1,{\mathfrak g})$-type and two-dimensional conformal field theories, which are observed in \cite{Beem:2013sza, Cordova:2015nma,Xie:2016evu}. 
In a previous paper \cite{Ito:2017ypt},  we have observed this 2d/4d-correspondence from the viewpoint of quantum Seiberg-Witten curve of the Argyres-Douglas theories. 
It would be interesting to study the relation between the quantum integrable models and superconformal field theories in four dimensions. 

It is also interesting to derive the T-/Y-system and the TBA equations for the $A_r^{(1)}$-type modified affine Toda field equations. 
The monodromy parameter $g$ leads to the non-trivial boundary conditions as observed in \cite{Saleur:2000bq,Lukyanov:2010rn,Maldacena:2010kp,Gao:2013dza,Ito:2016qzt}.
From the TBA equations, one also obtain the effective central charge at UV limit from, which should coincides with the one derived from the NLIEs. 
For $A_1^{(1)}$ case with $M$ being positive integer, we can confirm that both calculations agree with 
each other. But for higher rank case, it is a non-trivial problem, which should be investigated.

\subsection*{Acknowledgements}
We would like to thank S. L. Lukyanov, J. Suzuki and R. Tateo for useful discussions. 
The work of HS is supported in part by JSPS Research Fellowship for Young Scientists, from the Japan Ministry of Education, Culture, Sports, Science and Technology.
The work of KI is supported
in part by Grant-in-Aid for Scientific Research 
15K05043, 18K03643 and  16F16735 from Japan Society for the Promotion of Science (JSPS).

\appendix

\section{Connection with the NLIEs of $A_r$-type complex affine Toda model}\label{sec:massive-IM}
In section \ref{sec:NLIEs-Ar}, we derived the NLIEs from the Bethe ansatz equations. In this appendix, we rewrite these equations and show their relations with the ones obtained in \cite{ZinnJustin:1997at}. 
Under the identifies
\bee
%\delta_{m}&\leftrightarrow&0\mod2,\\
%-iZ_{m}&\leftrightarrow&\log a^{(m)},\quad 
e^{-iZ_{a}(\lambda)}&\leftrightarrow& a^{(a)}(\lambda),\quad m_{a}L\leftrightarrow 2b_{0}M_{a},\quad\\
%n&\leftrightarrow& r,\quad h=n+1,\\
\kappa&\leftrightarrow& \frac{\pi}{h}k,\quad \gamma=\frac{M\pi}{1+M},%\\
%X_{mt}&\leftrightarrow&\varphi_{mt},
\ee
(\ref{eq:DDV-Ar}) becomes
\bee\label{eq:NLIE-Zinn-Justin}
Z_{a}=m_{a}L\sinh\lambda-\pi\hat{\alpha}_{a}+\sum_{b=1}^{n}X_{ab}\ast \tilde{Q}_b
\ee
where 
\bee
\int d\lambda e^{i\kappa h\lambda/\pi}X_{ab}(\lambda)&=& \delta_{ab}-\frac{\sinh\frac{\pi\kappa}{\gamma}}{\sinh\kappa(\frac{\pi}{\gamma}-1)\cosh\kappa}\tilde{C}_{ab}^{-1}(\kappa)\\
\tilde{C}_{ab}^{-1}(\kappa)&=&\coth\kappa\frac{\sinh((n+1-a)\kappa)\sinh(b\kappa)}{\sinh((n+1)\kappa)}=C_{ab}^{-1}(k)\\
\tilde{Q}_{b}(x)&=&\frac{1}{i}\log\frac{1+e^{iZ_{b}(x+i0)}}{1+e^{-iZ_{b}(x-i0)}}.
\ee
It is easy to check $X_{ab}(\lambda)=\varphi_{ab}(\lambda)$.
(\ref{eq:NLIE-Zinn-Justin}) is the twisted NLIEs studied in \cite{ZinnJustin:1997at} without hole, special root or complex root. Note that (\ref{eq:NLIE-Zinn-Justin}) is derived for a class of integrable models associated with the quantum group $U_q(\hat{\mathfrak{g}})$. More precisely, (\ref{eq:NLIE-Zinn-Justin}) are the NLIEs for $A_r$-type complex affine Toda models.

\end{document}